\documentclass[twoside,11pt]{article}
\usepackage{epsfig}  
\usepackage{pst-plot}
\usepackage{amssymb}
\usepackage{bm}

\newcommand{\ome}{\omega_{e}}
\newcommand{\anu}{\bar\nu}
\newcommand{\omnu}{\omega_{\nu}}
\newcommand{\sla}{\! \not \!}
\newcommand{\be}{\begin{equation}}
\newcommand{\ee}{\end{equation}}
\newcommand{\bea}{\begin{eqnarray}}
\newcommand{\eea}{\end{eqnarray}}

\newcommand{\ep}{\varepsilon}
\def\Re{\mathop{\rm Re}}
\def\Im{\mathop{\rm Im}}

\newcommand {\vecq} {\bm q}
\newcommand {\vecx} {\bm x}

\newcommand {\vecp} {\bm p}

\newcommand {\vecsigma} {\bm \sigma}

\newcommand{\Tr}{{\rm Tr}}
\newcommand{\Cos}{{\rm cos~}}

\begin{document}
\title{Neutrino radiation from dense matter\footnote{Based in part on
the  lectures delivered at the Summer school 
``Dense Matter In Heavy Ion Collisions and Astrophysics", BLTP, Joint
Institute for Nuclear Research, Dubna, Russia. 
The complete lectures are available online at 
{\tt http//theor.jinr.ru/$\sim$dm2006/talks.html}   }}
\author{Armen~Sedrakian\\
 Institute for Theoretical Physics,\\
  T\"ubingen University,  D-72076 T\"ubingen, Germany
}

\maketitle

\begin{abstract}
This article provides a concise review of the problem 
of neutrino radiation from dense matter. The subjects 
addressed include quantum kinetic equations for neutrino 
transport, collision integrals describing neutrino 
radiation through charged and neutral current interactions, 
radiation rates from pair-correlated baryonic and color 
superconducting quark matter. 
\end{abstract}

\section{Introduction}\label{sec:INTRO}

After an initial phase of rapid (of the order of weeks to years) 
cooling from temperatures 
$T\sim 50$ MeV down to 0.1 MeV, a  neutron star's core settles in a thermal 
quasi-equilibrium state which evolves slowly over the time scales $10^3-10^4$ 
yr down to temperatures $T\sim 0.01$ MeV. The cooling rate of the star
during this period is determined by the processes of neutrino emission 
from dense matter, whereby the neutrinos, once produced, leave the star 
without further interactions. Understanding the cooling processes that
take place during this 
neutrino radiation era is crucial for the interpretation 
of the data on surface temperatures of neutron stars. While 
the long term features of the thermal 
evolution of neutron stars are insensitive to the initial rapid cooling stage,
the subsequent route in the temperature versus time diagram, which 
includes the late time ($t\ge 10^5$ yr) photoemission era,  strongly 
depends on the emissivity of matter during 
the neutrino cooling era.

This lecture is a concise introduction to 
the physics of neutrino radiation from 
dense nucleonic and quark matter in compact stars. It starts
with a classification of the reactions in Sec.~\ref{sec:class}, which is 
followed by a discussion of quantum kinetic equations for
neutrinos and neutrino emissivities in Sec.~\ref{sec:emiss}. 
In Sec.~\ref{sec:pol} examples are given of polarization tensors 
of superfluid nucleonic matter and color superconducting quark matter. 
We close by suggesting two exercises for students.

\subsection{Classification of the reactions}
\label{sec:class}
Historically, the weak reactions in neutron stars were classified 
within the quasiparticle description for fermions in matter:
each reaction is distinguished by the number of the participating 
quasiparticles and the  weak-interaction current.
The simplest neutrino emission processes that involve single fermionic
quasiparticle in the initial (final) state can be written as
\bea \label{URCA}
&&f_1 \to f_2 + e + \anu , \quad f_2 + e \to f_1 + \nu , \\ 
\label{BREMS}
&&f \to  f+ \nu +\anu,   \hspace{5cm} ({\rm forbidden})
\eea
where the first line is the charged current $\beta$-decay and its inverse,
with $f_1$ and $f_2$ being neutron and proton quasiparticles in nucleonic 
matter or $d$ and $u$ flavor quarks in deconfined quark matter; $f$ refers 
to a fermion. This process is known in astrophysics 
as the Urca processes~\cite{GAMOW}. 
The Urca reaction  is kinematically allowed in nucleonic matter under 
$\beta$-equilibrium  if the proton fraction is sufficiently 
large, $Y_p \ge 11-14\%$~\cite{Boguta81aLATTIMER_PRL}. In deconfined, 
chirally symmetric, and interacting quark matter at moderate densities 
the Urca processes is kinematically allowed for 
any asymmetry between $u$ and $d$ quarks~\cite{IWAMOTO}.
The second process - the neutral current neutrino pair 
bremsstrahlung, Eq.~(\ref{BREMS}),  is forbidden by the energy and 
momentum conservation, if one adopts the quasiparticle picture.
If, however, we choose to work with excitations that are characterized 
by finite widths, the reaction (\ref{BREMS}) is allowed~\cite{SD99}. 
The processes with two fermions in the initial (and final) states 
are the modified Urca and its inverse~\cite{MOD_URCA}
\bea
\label{MOD_URCA}
&&f_1 +f_1 \to f_1+ f_2 + e + \anu , \quad f_1+f_2 \to f_2+ f_2 + e + \anu, \\ 
\label{MOD_URCA_INV}
&&f_1 + f_2 + e \to f_1+ f_1+ \nu , \quad f_2 + f_2 + e \to f_2 + f_1 + \nu,
\eea
and the modified bremsstrahlung process
\bea
\label{MOD_BREMS}
&&f+f \to f + f + \nu +\anu .
\eea
The modified processes are characterized by a spectator baryon that 
guarantees energy and momentum conservation in baryonic matter. 
In quark matter these processes are subdominant 
due to the extra phase space required by the spectator quarks. Indeed
each extra fermion in the initial and final state introduces a 
small factor $T/E_F\ll 1$, where $E_F$  is the Fermi  energy.
The general arguments above apply to the reactions in
quark matter featuring strange quarks and in the hypernuclear
matter, where the kinematical constraints are less restrictive
than in purely nucleonic matter~\cite{DUNCAN,PRAKASH_H}.

Due to the attractive component of the strong interaction nucleons and 
quarks form Cooper pairs at sufficiently 
low temperatures (for up-to-date reviews on nuclear and quark 
superconductivity see Refs.~\cite{NU_SUP,QUARK_SUP}). The formation 
of Cooper condensates lifts the constraint on the neutral current 
one-body processes in nucleonic~\cite{PAIR_BRAKING,VOSK87}
and quark matter~\cite{JAIKUMAR}, thus leading to the reaction 
\be \label{CPBF}
\{ff\} \to f+f+\nu_f+\anu_f, \quad f+f\to \{ff\} + \nu_f + \anu_f,
\ee
where $\{ff\}$  refers to a Cooper pair, $f+f$ to two quasiparticle
excitations. These processes - termed Cooper pair breaking and 
formation (CPBF) reactions~\cite{SCHAAB}  
- are efficient in the temperature domain 
$T^*\le T\le T_c$,  where $T_c$ is the critical temperature of 
superfluid phase transition and $T^*\sim 0.2~ T_c$; they are suppressed 
asymptotically at low temperatures as  ${\rm exp}(-\Delta(0)/T)$, 
where $\Delta(0)$ is the zero-temperature pairing gap. The temperature
domain above matches firmly with characteristic temperatures in the 
neutrino cooling era ($T_c \sim$ MeV for nucleonic matter). Thus, the 
CPBF processes are an important ingredient of the cooling of at least
the nucleonic matter. The case of quark matter is less clear: the critical 
temperature of pairing of quarks in the dominant pairing channels 
could be as large as 50 MeV; however smaller, $\sim$ keV, gaps were predicted 
for some combinations of quantum numbers, and the associated critical 
temperatures lie within the relevant temperature range~\cite{QUARK_SUP}.
The neutral current processes (\ref{CPBF}) 
induced by the superfluidity  have their charged current 
counterparts~\cite{SEDRAKIAN05,JRS}. While the former vanish, when the 
temperature approaches the critical temperature of superfluid phase 
transition, the emissivity of the latter process approaches the value 
of the corresponding Urca process.

\section{Quantum kinetics of neutrinos in matter}
\label{sec:emiss}
Among the methods that are used to compute the rates of neutrino 
production in dense matter those that use the language of many-body 
theory are particularly suited, as the whole approach can be organized
in a systematic way, that is  consistent with the treatment of related 
problems of the equation of state, specific heat of matter, pairing fields,
etc. In particular, the formulations based on the real-time Green's 
functions (RTG) technique allow for treatments of non-equilibrium 
processes, including
situations far from equilibrium. The RTG technique 
was applied to compute the neutrino emissivities for several reactions 
in nucleonic matter by Voskresensky and Senatorov~\cite{VOSK87}. 
In their approach the rates are computed from the $S$-matrix with the 
help of the optical theorem. Alternatively, 
the neutrino emission rates can be derived directly from a quantum 
kinetic equation for neutrinos, whereby the collision integrals 
are expressed in terms of neutrino self-energies~\cite{SD99,SEDRAKIAN99b}. 
Below, the latter method will be illustrated on a few examples.

\subsection{\it Transport equations for neutrinos}

We wish to write down a transport equation for neutrinos in a general 
form involving only Green's functions and self-energies. One way of 
doing this is to start with the Dyson equation for neutrinos written 
on a real time contour. At the first order in the gradient expansion, 
and upon taking the quasiparticle limit in neutrino propagators, one
finds~\cite{SEDRAKIAN99b}
\bea\label{QPA_TRANS}
  i\left\{\Re S^{-1}(q,x),S_0^{>,<}(q,x)\right\}_{P.B.}
   =S^{>,<}(q,x)\Omega^{>,<}(q,x)+\Omega^{>,<}(q,x)S^{>,<}(q,x),
\eea
where $S^{>,<}(q,x)$ and $\Omega^{>,<}(q,x)$ are the neutrino propagators
and self-energies, $q = (q_0,\vecq)$ and $x= (t,\vecx)$ are the four-momentum 
and space-time coordinates, $\{\dots\}_{P.B.}$ is the four-dimensional 
Poisson bracket; the symbols $>,< $ refer to the positioning 
of the time arguments 
of the two-point functions $S$ and $\Omega$ on the real-time contour; 
$\Re S^{-1}(q,x)$ is the inverse of the retarded Green's function.
The l. h. side of Eq.~(\ref{QPA_TRANS}) corresponds to the drift term of 
the Boltzmann equation (hereafter BE), while the r. h. is the collision 
integral. The on-mass-shell neutrino propagator is related to the single-time
distribution functions (Wigner functions) of neutrinos and anti-neutrinos,
$f_{\nu}(q,x)$ and $f_{\bar\nu}(q,x)$, via the ansatz
\bea
S_0^<(q,x)
&=& \frac{i\pi\sla q}{\omnu(\vecq)}
    \Big[ \delta\left(q_0-\omnu(\vecq)\right)f_{\nu}(q, x)
-\delta\left(q_0+\omnu(\vecq)\right) 
\left(1-f_{\bar \nu}(-q,x)\right) \Big],
\eea
where $\omnu(\vecq)=\vert \vecq\vert$ is the on-mass-shell
neutrino/anti-neutrino energy.  Note that the ansatz
includes {\it simultaneously} the neutrino particle states
$\propto f_{\nu}(q, x)$ and anti-neutrino hole states 
$\propto 1-f_{\bar \nu}(-q,x)$.

Upon applying the  trace operation (in the space of Dirac matrices) 
on both sides of the transport equation  (\ref{QPA_TRANS}) and integrating
out the off-shell energies on the l. h. side, one obtains 
a single time BE for neutrinos 
\bea\label{BE_NU}
& & \left[\partial_t + \vec \partial_q\,\omnu (\vecq) \vec\partial_x
\right] f_{\nu}(\vecq,x) =
\int_{0}^\infty \frac{dq_0}{2\pi} {\rm Tr} \left[\Omega^<(q,x)S_0^>(q,x)
-\Omega^>(q,x)S_0^<(q,x)\right];
\eea
a similar equation follows for the  anti-neutrinos if one integrates
in Eq. (\ref{QPA_TRANS})
over the range $[-\infty , 0]$.

\subsection{\it Collision integrals}
\label{COLLISION}
\begin{figure}[tb]
\begin{center}
\epsfig{figure=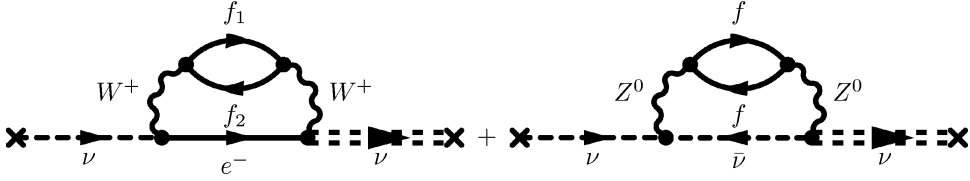,width=12.8cm,angle=0}
\begin{minipage}[t]{17.5 cm}
\caption{ The neutrino self-energy in the case of charged (left) 
and neutral (right) current interaction mediated by $W^+$ and $Z^0$ 
gauge bosons, respectively; the baryon/quark propagators are 
labeled as $f_i$, $i=1,2$; that of neutrino, anti-neutrino and electrons
as $\nu$, $\anu$ and $e^-$. 
Note that the $\nu$ and $\anu$ propagators are shown for 
illustration and should not be included in the evaluation of the diagram.
}\label{fig:dyson_nu}
\end{minipage}
\end{center}
\end{figure}

Leading order contributions to the 
neutrino radiation rates arise from second order Born diagrams
when the neutrino self-energies $\Omega^{>,<}(q)$ are expanded with respect 
to the weak coupling constant. The diagrams contributing to the charged 
and neutral current processes are shown in Fig.~\ref{fig:dyson_nu}. 
The corresponding neutrino self-energies are given by 
\bea
-i\Omega^{>,<}(q_1,x) &=& \int \frac{d^4 q}{(2\pi)^4}
\frac{d^4 q_2}{(2\pi)^4}(2\pi)^4 \delta^4(q_1 - q_2 - q)
i\Gamma_{ q}^{\mu}\, iS_0^{<}(q_2,x) i\Gamma_{ q}^{\dagger\, \lambda}
i \Pi^{>,<}_{\mu\lambda}(q,x),
\eea
where $\Pi^{>,<}_{\mu\lambda}(q)$ refer to the polarization tensors,
$\Gamma_{ q}^{\mu}$ is the weak interaction vertex to be specified below. 
The central problem of the theory is to compute the polarization 
tensors of nucleonic or quark matter. 

To obtain the emissivity through, e. g., a charged current process, 
we compute from the BE the change in the energy per unit volume and 
time due to the change in the anti-neutrino distribution
\bea\label{EMISSIVITY2}
\epsilon_{\anu}&=& \frac{d}{dt}\int\!\frac{d^3q}{(2\pi)^3}
f_{\anu}(\vecq)\omnu(\vecq) =               
- 2\left( \frac{\tilde G}{\sqrt{2}}\right)^2
\int\!\frac{d^3q_1}{(2\pi)^32 \ome(\vecq_1)}
\int\!\frac{d^3 q_2}{(2\pi)^3 2\omnu(\vecq_2)}
\int\! d^4 q \,\delta (\vecq_1 + \vecq_2 -  \vecq)
\nonumber\\
&&\delta(\ome+\omnu-q_{0})\, \omnu(\vecq_2)
 g_B(q_0)\left[1-f_{\anu}(\omega_{\anu})\right]\left[1-f_{e}(\ome)\right]
 \Lambda^{\mu\zeta}(q_1,q_2)\Im\,\Pi^R_{\mu\zeta}(q),
\eea
where $\tilde G = G\Cos \theta_C$,  $G$ is the weak coupling constant,
 $\theta_C$ is the Cabibbo angle ($\Cos \theta_C = 0.973$) and
$\Lambda^{\mu\zeta}(q_1,q_2) = {\rm Tr}\left[\gamma^{\mu}
(1 - \gamma^5)\sla q_1\gamma^{\zeta}(1-\gamma^5)\sla q_2\right]$.
The symbol $\Im$ refers to the imaginary part of the polarization 
tensor's resolvent. Here we used the relation $\Pi_{\mu\zeta}^{<}(q)
=\Pi_{\zeta\mu}^{>}(-q) = 2i g_B(q_0) {\Im}\,
\Pi_{\mu\zeta}^R(q)$, where $g_B(q_0)$ is the Bose distribution
function and $\Pi^R_{\mu\zeta}(q)$ is the retarded component
of the polarization tensor. In equilibrium,  $f_{\anu/e}(\omega_{\anu/e})$ 
reduce to Fermi-distribution functions for anti-neutrinos 
and electrons. Since the anti-neutrinos leave the star without 
interactions, there is no thermal population of anti-neutrinos, i.~e. 
$f_{\anu}(\omega_{\anu}) \ll 1$
and can be neglected. The neutrino emissivity for the case of neutral current 
processes can be obtained in a similar way~\cite{SD99,SEDRAKIAN99b}.

\section{\it Polarization tensors of dense matter: Examples}
\label{sec:pol} 

It is instructive to study the polarization tensors describing 
charged and neutral current processes first at the single loop level. 
Descriptions that are consistent with the 
conservation laws  and Ward identities require vertex 
corrections to the one-loop results, 
which we shall address later on. We shall now switch to 
the equilibrium finite temperature techniques of Matsubara Green's 
functions thus treating the nucleonic/quark matter in thermal 
equilibrium.

\subsection{\it Direct Urca process in baryonic matter}

Since the temperature of dense matter core during the neutrino 
cooling era is well below the critical temperatures of pairing 
in baryonic and quark matter, pairing correlations should be included
in the computation of polarization tensors. The non-relativistic 
Matsubara propagators that incorporate the pairing correlations are given by 
\bea \label{P5}
\hat G^M_{\alpha\alpha'}(ip_n,\vecp) 
&=& \delta_{\sigma\sigma'}\delta_{\tau\tau'}\left(
\frac{u_p^2}{ip_n-\ep_p} +\frac{v_p^2}{ip_n+\ep_p} \right) = 
\delta_{\sigma\sigma'}\delta_{\tau\tau'}G^M(ip_n,\vecp),\\
     \label{P6} 
\hat F_{\alpha\alpha'}(ip_n,\vecp) 
&=& - i\sigma_y\delta_{\tau\tau'}
              u_pv_p\left(\frac{1}{ip_n-\ep_p}-\frac{1}{ip_n+\ep_p}\right),
\eea
where $p_n = (2n+1)\pi T$  is the fermionic Matsubara frequency,
$\sigma$ and $\tau$ refer to spin and isospin, $\sigma_y$ is 
the $y$-component of the Pauli-matrix, $u_p^2 = (1/2)(1+\xi_p/\ep_p)$ 
and $v_p^2 = 1-u_p^2$ are the Bogolyubov amplitudes and 
$\ep_p= \sqrt{\xi_p^2+\Delta_p^2}$ is the quasiparticle spectrum, 
where $\xi_p = p^2/2m^* -\mu$ is the spectrum in the unpaired 
state, with $m^*$ and $\mu$ being the effective mass and chemical 
potential. Here $\Delta_p$ is the anomalous self-energy (gap function).
The propagators above are written for the case of $S$-wave 
neutron or proton proton pairing in isospin-1, spin-0 state. 
Note that at high densities the neutron fluid is paired in a $P$ wave
(this is not the case for protons because of their low abundance).
Consider now the case of direct Urca process
involving nucleons (Fig.~\ref{fig:dyson_nu}, left diagram). The 
Matsubara polarization tensor is then given by 
\be\label{eq:pol_tensor} 
i\Pi_{\mu\nu}^M(iq,\vecq) = \sum_{ip}\int\frac{d^3p}{(2\pi)^3}
\Tr \left[\Gamma_{\mu} G_{\tau=1/2}^M(ip,\vecp)\Gamma_{\nu} 
G_{\tau=-1/2}^M(ip+iq,\vecp+\vecq)\right],
\ee
where the charged current weak interaction vertices are 
$\Gamma_{\mu} =  \gamma_{\mu}(1- g_A\gamma^5)$, with 
$g_A = 1.26$ being the axial coupling constant.   
Upon performing the Matsubara sums and analytical continuation 
$(iq \to \omega + i\delta)$ we obtain the retarded polarization tensor
 \bea \label{POL_TENS_URCA2}
 \Pi_{V/A}^{R} (iq,\vecq)  &=& 2\int\frac{d^3p}{(2\pi)^3}
 \Biggl\{
\left(\frac{u_p^2u_{p+q}^2}{\omega+\ep_p-\ep_{p+q}+i\delta}
-\frac{v_p^2v_{p+q}^2}{\omega-\ep_p+\ep_{p+q}+i\delta}
\right)\left[f(\ep_p)-f(\ep_{p+q})\right]\nonumber\\
&&\hspace{-1.5cm}+
u_p^2v_{p+q}^2\left(\frac{1}{\omega-\ep_p-\ep_{p+q}+i\delta}
-\frac{1}{\omega+\ep_p+\ep_{p+q}+i\delta}
\right)\left[1-f(\ep_p)-f(\ep_{p+q})\right]\Biggr\},
 \eea
where the vector/axial-vector polarization tensors $\Pi^R_{V/A}$ are
the components proportional to $1$ and $g_A^2$, respectively.
The first two terms in Eq.~(\ref{POL_TENS_URCA2})
correspond to  excitations of a particle-hole pair while the
last two to excitation of particle-particle and hole-hole
pairs. The last term does not contribute to the neutrino
radiation rate $(\omega > 0)$. We identify the first two
terms as the scattering ($SC$) terms, while the third
term as the pair-braking ($PB$) term. 
Upon evaluating the phase space integrals,
the neutrino emissivity is written as $\epsilon_{\anu}= \epsilon_0^{\rm Urca} 
J,$ where
\bea\label{SUP_EMISSIVITY4}
\epsilon_0^{\rm Urca} 
=(1+3g_A^2)\frac{3\tilde G^2 m_{n}^*m_{p}^*p_{Fe}T^6}{2\pi^5},
\quad
J = -\frac{1}{6}\int_{-\infty}^{\infty}\!dy~g_B(y)
\left[I^{SC}(y)+ I^{PB}(y)\right]\int_{0}^{\infty} dz z^3
f_e(z-y),
\eea
where $p_{Fe}$ is the Fermi-momentum of the electrons 
and $y = \beta\omega$; the integrals $I^{SC}(y)$ 
and $I^{PB}(y)$  are given in Refs.~\cite{SEDRAKIAN05}.
In the unpaired state ($u_p\to 1$ and $v_p\to 0$) only the scattering 
contribution survives; upon integrating we obtain
\bea\label{URCA_EMISSIVITY} 
I^{SC}(y) = {\rm ln}\Bigg\vert
\frac{1+{\rm exp}\left[-\beta\xi\right]}
{1+{\rm exp}\left[-\beta(\xi +\omega)\right]}\Bigg\vert ,
\eea
where $\xi = \tilde p^2/2m^* - \mu_p$ and $\tilde p = (m^*/q) (\omega -
\mu_p +\mu_n -q^2/2m^*)$; here the momentum transfer $q = p_{Fe}$, 
$\mu_n$ and $\mu_p$ are the chemical potentials of neutrons 
and protons, and we assumed for simplicity 
that their effective masses are equal. In the zero temperature limit 
$ I^{SC}(y) = y\theta (-\beta\xi)$, the integrals in 
Eq.~(\ref{SUP_EMISSIVITY4}) can be performed analytically and one 
recovers the zero-temperature result of 
Lattimer et al.~\cite{Boguta81aLATTIMER_PRL}.
The zero temperature $\theta$-function can be rewritten as 
$\theta (p_{Fe} + p_{Fp} -p_{Fn})$~\cite{Boguta81aLATTIMER_PRL} which tells 
us that the ``triangle inequality" $p_{Fe} + p_{Fp} \ge p_{Fn}$ must 
be obeyed by the Fermi-momenta of the particles for the Urca process 
to operate. 

While the one-loop approximation provides a useful starting point, 
the complete treatment of the problem when the particle-hole interaction 
is not small, i.~e. can not be treated as a perturbation, requires 
summation of infinite series of particle-hole loops. This is certainly 
the case in nuclear matter, where the Landau parameters are $O(1)$.
\begin{figure}[t]
\begin{center}
\includegraphics[height=1.7cm,width=15cm]{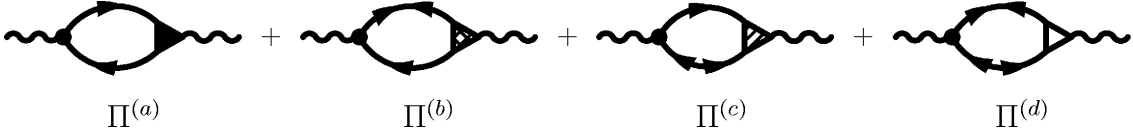}
\end{center}
\caption[]
{The sum of polarization tensors that  contribute to the neutrino 
emission rate.  The single arrowed solid line corresponds to the normal 
propagator (\ref{P5}), while the double arrowed line to the anomalous
propagator (\ref{P6}). The contributions form $\Pi^{(b)}(q)$,
$\Pi^{(c)}(q)$, and $\Pi^{(d)}(q)$ diagrams vanish at one-loop for the 
Urca process; the $\Pi^{(d)}(q)$ diagram is finite for charge neutral 
interactions at one-loop.
}
\label{fig:pol_sup}
\end{figure}
Figure~\ref{fig:pol_sup} 
shows the four distinct diagrams in the case where the loops 
are summed up to all orders. Next subsection shows how to improve 
on the one-loop result using as an example the neutral current interactions.


\subsection{\it Neutral current neutrino pair-bremsstrahlung}

The polarization tensor describing the neutral current interactions 
in baryonic matter is given by 
\be \label{sec2:eq1}
i\Pi_{\mu\nu}(q) = \sum_{ip}\int\frac{d^3p}{(2\pi)^3}
\Tr \left[ \Gamma_{\nu}G(ip,\vecp)\Gamma_{\mu}G(ip+iq,\vecp+\vecq) 
+ \Gamma_{\nu}\hat F(ip,\vecp)\Gamma_{\mu}\hat 
F^{\dagger} (ip+iq,\vecp+\vecq) \right],
\ee
where the neutral current vertices are 
$\Gamma_{\mu} =   \gamma_{\mu}(c_V- c_A\gamma^5)$.
Performing the Matsubara sums we obtain for the vector and axial-vector 
contributions in this case
\bea \label{sec2:eq2}
\Pi^{V/A}(q) &=& \sum_{\sigma\vecp} 
\left[f(\ep_p) -f(\ep_k)\right] \left(
\frac{A_{\mp}}{iq+\ep_p-\ep_k} -\frac{B_{\mp}}
{iq-\ep_p+\ep_k}\right)\nonumber\\
&+&\sum_{\sigma\vecp} \left[f(-\ep_p) -f(\ep_k)\right] \left(
\frac{C_{\mp}}{iq-\ep_k-\ep_p}-\frac{D_{\mp}}{iq+\ep_p+\ep_k}
\right),
\eea
where $k = p +q$, $A_{\mp} = u_p^2u_k^2 \mp h$, 
      $B_{\mp} = v_p^2v_k^2 \mp h$,
      $C_{\mp} = u_k^2v_p^2 \pm h$,
      $D_{\mp} = u_p^2v_k^2 \pm h$, 
$h = u_pu_kv_pv_k$. The first line in Eq.~(\ref{sec2:eq2}) 
corresponds to the process of scattering
where a quasiparticle is promoted out of the condensate into 
an excited state, or inversely, an excitation merges with 
the condensate.  The corresponding piece of the response 
function $\Im\Pi^{V/A}(q)$ vanishes for small momentum transfers. 
The second line in Eq. (\ref{sec2:eq2}) describes the process 
of pair-breaking and recombination, i.~e., excitation of pairs 
of quasiparticles out of the condensate, and inversely, 
restoration of a pair within the condensate. Since we are interested
in the emission process we shall keep only the terms that 
do not vanish for $\omega > 0$; then, the pair-braking contribution 
is given by the term $\propto C_{\pm}$.
This contribution to the polarization tensor can be 
evaluated analytically in the limit $\vecq\to 0$ and 
the case $\Delta\neq \Delta(p)$ and is given by 
\bea\label{sec2:eq5}
\Im\,\Pi^{V}(q) &=& - 2\pi \nu(p_F)
g(\omega)^{-1} f\left(\frac{\omega}{2}\right)^2
\left(\frac{\Delta^2}{\omega^2}\right)
\frac{\omega}{\sqrt{\omega^2-4\Delta^2}}\theta(\omega-2\Delta),\\
\Im\,\Pi^{A}(q) &\simeq & 0 + O\left(\frac{v_F^2}{c^2}\right),
\eea
where  $\nu(p_F) = m^* p_F/2\pi^2$ is the density of states ($\hbar = 1$)
and $\theta$ is the Heaviside step function; the explicit 
form of the $O\left({v_F^2}/{c^2}\right)$ contribution to the 
axial current response is given by Flowers et al in Ref.~\cite{PAIR_BRAKING}. 

Upon substituting Eq.~(\ref{sec2:eq5}) in the neutral current 
analog of Eq.~(\ref{EMISSIVITY2}) 
and carrying out the phase-space integrals we obtain the
emissivity per neutrino flavor~\cite{PAIR_BRAKING}
\be \label{EMISSIVITY5}
\epsilon_{\nu\anu} 
= \frac{G^2\, c_V^2}{240\pi^3}~~\nu(p_F)~T^7~ 
I_1(\zeta)\equiv \epsilon_0^{\rm brems.}~I_1(\zeta),
\ee
where $\zeta = 2\Delta(T)/T$ and
\be \label{INT2}
I_1(\zeta) = \zeta^7 \int_0^{\infty}\!\! d\phi ~({\rm cosh}\,\phi)^5 
~f\left(\frac{\zeta}{2}\, {\rm cosh}\,\phi\right)^2.
\ee
Note that the rate (\ref{EMISSIVITY5}) scales as $\zeta^7$ and,
consequently, it is sensitive to the  magnitude of the pairing gap.
Because of the substantial density dependence 
of the gap, the emissivity (\ref{EMISSIVITY5}) varies strongly across
the stellar interior.

\begin{figure}[t]
\begin{center}
\includegraphics[height=4.5cm,width=15cm]{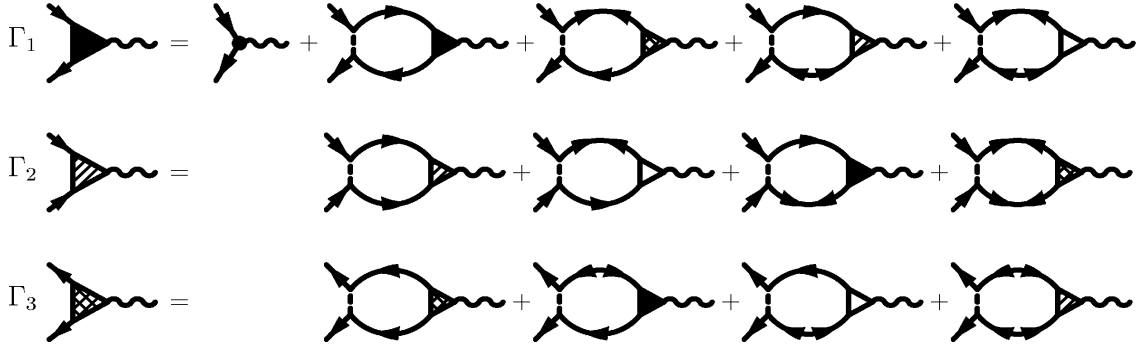}
\end{center}
\caption[]
{
Coupled integral equations for the effective weak vertices
in superfluid baryonic matter. The ``normal'' $\Gamma_1$ vertex 
(full triangle) and two ``anomalous'' vertices $\Gamma_2$ 
(hatched) and $\Gamma_3$ (shaded triangle) are shown 
explicitly, the fourth vertex (empty triangle) is obtained by 
interchanging the particle and hole lines in the first line.
The anomalous vertices vanish in the normal state.
}
\label{fig:vertex}
\end{figure}

In nuclear and neutron matter problem the particle-hole interactions 
are not small and cannot be treated in the perturbation theory.
The resummation of particle-hole diagrams in a superfluid 
leads to coupled integral equations shown in Fig.~\ref{fig:vertex}. 
In the non-relativistic limit the driving terms in the vector and 
axial-vector channels correspond to the scalar and spinor perturbations, 
i.~e. the bare vector and axial-vector vertices are $\Gamma^V = 1$ 
and $\Gamma^A = \vecsigma$. The topologically non-equivalent polarization
tensors and the associated vertices are shown in Figs.~\ref{fig:pol_sup} and
~\ref{fig:vertex}.

Including the vertex corrections modifies the one-loop result to 
(Sedrakian et al. in ref.~\cite{PAIR_BRAKING})
\be \label{EMISSIVITY3}
\epsilon_{\nu\anu} =  \epsilon_0~(I_2+I_3),
\ee
where
\bea\label{eq:integrals}
I_2 &=&  -\frac{1}{\pi\nu(p_F) T^7}
\int_{0}^{\infty}\!\! 
d\omega ~\omega^6 g(\omega)\sum_{i=1}^3 \left[{\rm Im}\Pi_i(\omega) 
~{\rm Re} \Gamma_i(\omega)\right],\\
I_3 &=&  -\frac{1}{\pi \nu(p_F) T^7}
\int_{0}^{\infty}\!\! d\omega ~\omega^6 g(\omega)
\sum_{i=1}^3 \left[{\rm Re}\Pi_i(\omega) 
~{\rm Im} \Gamma_i(\omega)\right],
\eea
where $\Pi_1 = \Pi^{(a)}- \Pi^{(d)}$, $\Pi_2 =\Pi^{(b)}$ 
and $\Pi_3 =\Pi^{(c)}$. For $T\to T_c$ the  rates vanish, 
consistent with the observation that the pair bremsstrahlung 
is absent in normal matter for on-shell (non-interacting) 
baryons. At small $T\le 0.3 T_c$ the rates are suppressed 
exponentially as ${\rm exp}(-\Delta/T)$.

\subsection{\it Direct Urca process in color superconducting matter}
Although quark matter in compact stars and its superfluidity were suggested
more than three decades ago, these topics have received much attention in 
recent years after the models, which were designed to describe the chiral 
phase transition in dense matter, were applied to the problem of quark 
superconductivity~(the current state of the art is reflected in the 
reviews~\cite{QUARK_SUP}). 
At moderate densities relevant to compact stars the quark 
matter is in the non-perturbative regime and one has to rely on effective
models that capture (at least some) features predicted by the QCD 
(chiral symmetry breaking, confinement, etc.)
The ground state of superconducting quark matter under $\beta$-equilibrium 
is not known; one significant problem is that under the stress caused by 
$\beta$-equilibrium and/or the strange quark mass, 
the Fermi-surfaces of up ($u$) and down ($d$) quarks are shifted apart, 
and the resulting pairing patterns differ from the ones predicted by 
the BCS theory. Color and flavor degrees of freedom are responsible for 
the multitude of possible pairing patterns. 

In contrast to the case of nucleonic matter the Urca process is 
permitted in {\it interacting} quark matter for any asymmetry between 
$u$ and $d$ quarks. The emissivity to first order in the strong coupling
constant $\alpha_c$ is given by~\cite{IWAMOTO}
\begin{equation}
\label{redemiss}
\epsilon_{0}=\frac{914}{315}~\tilde G_F^2~\alpha_c~\mu_d~\mu_u~\mu_e~T^6,
\end{equation}
where $\mu_i$, $i=d,\, u,\, e\, $ are the chemical potentials of down and
up quarks and electrons. If non-superconducting quark matter is present 
in the core of a neutron star, the star will cool very rapidly to 
temperatures well below the observational threshold. 

\begin{figure}[t]
\begin{center}
\includegraphics[width=0.45\textwidth]{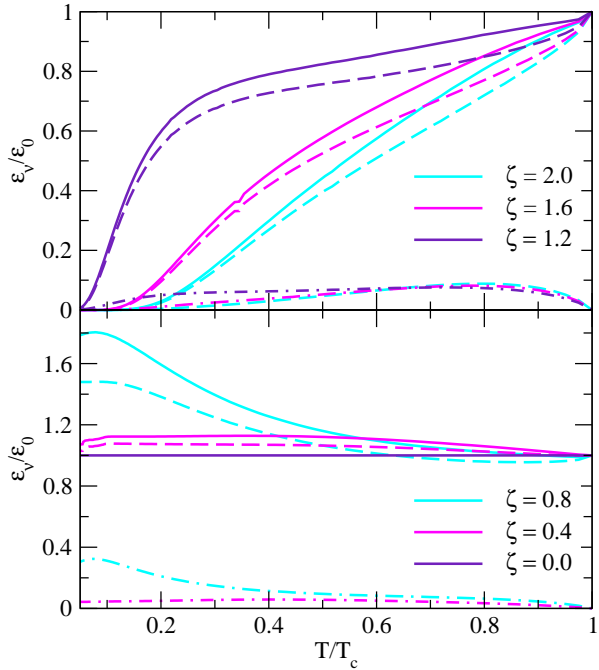}
\end{center}
\caption{Temperature dependence of the neutrino emissivity
 normalized to its value at the critical temperature
for a range of values of $\zeta = \Delta(0)/\delta\mu$; the upper panel 
corresponds to the gapped regime ($\zeta>1$) while the lower panel to the 
gapless regime ($\zeta <1$). The {\it dashed} and {\it dashed-dotted} 
lines are the  normal and anomalous contributions; 
the {\it solid} line is their sum.
}
\label{fig:qurca_fig2}
\end{figure}

What are the effects of superfluidity on the cooling rate of such a star? Let 
us consider a specific model~\cite{JRS}, where the pairing is in the so-called 
2SC phase, i.~e. quark pairing is characterized by the order parameter 
$\Delta \propto \langle
\psi^T(x)C\gamma_5\tau_2\lambda_2\psi(x)\rangle$, where
$\tau_2$ is the Pauli matrix in the isospin state, $\lambda_2$
is the Gell-Mann matrix in the color space, $C = i\gamma^2\gamma^0$ 
is the matrix of charge conjugation. The minimal effective Lagrangian 
describing the pairing is given by 
\be\label{eq1}
{\cal L}_{\rm eff} = \bar \psi (x)
(i\gamma^{\mu}\partial_{\mu})\psi(x)
+G_1(\psi^TC\gamma_5\tau_2\lambda_A\psi(x))^{\dagger}
(\psi^TC\gamma_5\tau_2\lambda_A\psi(x)),
\ee
where $G_1$ is the attractive pairing interaction. The Lagrangian is
minimal in the sense that apart from the pairing interaction it includes 
only the kinetic term for massless quarks; other channels of interaction, 
such as the repulsive components which would lead to the renormalizations 
of the single-particle spectra of quarks are omitted.
The normal and anomalous propagators of quarks of flavor $f$ are
\be 
S_{f = u,d}= i\delta_{ab}\frac{\Lambda^+(p)}{(p_0+\delta\mu)^2-\ep_p^2}
(\sla p - \mu_f\gamma_0),\quad \quad 
F(p) = -i\epsilon_{ab3}\epsilon_{fg}\Delta 
\frac{\Lambda^+(p)}{(p_0+\delta\mu)^2-\ep_p^2}\gamma_5C ,
\ee
where $\ep_p = \sqrt{(p-\mu)^2+\Delta^2}$,  $\delta\mu = (\mu_d-\mu_u)/2$
and $\mu = (\mu_d+\mu_u)/2$, $\Lambda^+$ is the projector to the positive
energy state. The emissivity at one-loop can be obtained by evaluating the 
sum 
\begin{figure}[tb]
\begin{center}
\epsfig{figure=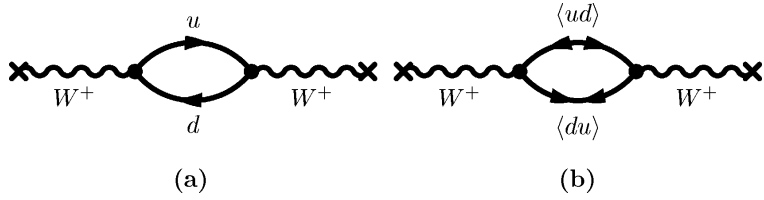,width=10.cm,angle=0}
\begin{minipage}[t]{17.5 cm}
\caption{ One-loop $W$-polarization tensor: (a) normal piece, which represents
  the sole contribution above $T_c$; and (b) ``anomalous'' piece, which is
  proportional to $\Delta^2(T)$ and vanishes at $T_c$. 
}\label{fig:loops_urca}
\end{minipage}
\end{center}
\end{figure}
of diagrams in Fig.~\ref{fig:loops_urca}, which leads to the polarization 
tensor
\be \label{eq:quark_pol}
\Pi_{\mu\lambda}(q) = -i\int\frac{d^4p}{(2\pi)^4} {\rm Tr}
\left[(\Gamma_-)_{\mu}S(p)(\Gamma_+)_{\lambda}S(p+q)
+(\Gamma_-)_{\mu}F(p)(\Gamma_+)_{\lambda}F(p+q)\right],
\ee
where $\Gamma_{\pm}(q) = \tilde G
\gamma_{\mu}(1-\gamma_5) \otimes\tau_{\pm}$
and $\tau_{\pm}=(\tau_1\pm\tau_2)/2$ are flavor-raising and lowering
operators. The result for the emissivity depends on whether the parameter
$\zeta = \Delta/\delta\mu$ is larger or smaller than unity~\cite{JRS}. 
For $\zeta > 1$ all the particle modes are ``gapped'',
therefore, as the temperature is lowered, the emissivity is suppressed
(for  asymptotically low temperatures exponentially). When $\zeta < 1 $
there are gapless modes in the quasiparticle spectrum; this implies 
that the neutrino production is not affected by color superconductivity
for these modes. As a result, the superconducting quark matter cools 
at a rate comparable to the unpaired matter. Fig.~\ref{fig:qurca_fig2}
illustrates these two distinct cases.

The neutrino emissivity of color-superconducting matter has been studied 
for alternative realizations of the ground state matter: 
one such realization is the spin-1 color 
superconductivity~\cite{SPIN1}.
Since the condensate in this case breaks the rotational symmetry 
the neutrino emission turns out to be anisotropic for some 
choices of the order parameter~\cite{SSW}. 
Another realization is the crystalline color superconductivity,
which is characterized by spatially modulated gap parameter. 
The neutrino radiation rates from crystalline color superconducting
matter and  cooling of compact stars featuring such a phase is discussed in 
Refs.~\cite{ANGLANI}.

\section{\it Suggested exercises}

\begin{enumerate}              
\item Derive the emissivity of the direct Urca process $n\to p + e+ \anu$ 
in unpaired matter by using the Fermi Golden rule and following 
the similar derivation  of the emissivity of the modified Urca process
$n+n\to n+ p + e+ \anu$ in Ref.~\cite{ST}. 
Rederive the emissivity  in unpaired matter 
by setting $u_p = 1$ and $v_p = 0$ in 
Eq.~(\ref{POL_TENS_URCA2}) and $\Lambda^{\mu\lambda} {\rm Im}
\Pi^R_{\mu\lambda} \simeq 8\omega_e\omega_{\nu} (
{\rm Im}\Pi_V + 3g_A^2{\rm Im}\Pi_A)$
in Eq.~(\ref{EMISSIVITY2}).

\item Derive the emissivity of unpaired quark matter featuring
$u$ and $d$ quarks through the direct Urca process $d\to u + e+\anu$
by using the Fermi Gold rule in the case where the 
quarks interact to leading order in $\alpha_s$~\cite{IWAMOTO}.
Repeat the calculation by starting from the polarization tensor 
(\ref{eq:quark_pol}) with $\Delta = 0$ (see Ref.~\cite{JRS} for the 
case $\Delta\neq0$). Next assume that quarks are 
non-interacting but massive. Repeat the calculations in 
this case and compare to the result of Ref.~\cite{IWAMOTO}. 
\end{enumerate}

\end{document}